\documentclass[aps,prb,reprint]{revtex4-2}
\usepackage{graphicx}
\usepackage{xspace}
\usepackage{hyperref}

\draft 

\def\nhvo{NH$_4^+$-V$_2$O$_5$\xspace}

\begin{document}

\title{Chains of magnetic ions in NH$_4^+$-intercalated vanadium pentoxide}

\author{Anna~A.~Anisimova}
\affiliation{M.N. Mikheev Institute of Metal Physics of Ural Branch of Russian Academy of Sciences, 18 S. Kovalevskaya St., Yekaterinburg 620137, Russia}

\author{Yaroslav~M.~Plotnikov}
\affiliation{M.N. Mikheev Institute of Metal Physics of Ural Branch of Russian Academy of Sciences, 18 S. Kovalevskaya St., Yekaterinburg 620137, Russia}

\author{Dmitry~M.~Korotin}
\email[]{dmitry@korotin.name}
\affiliation{M.N. Mikheev Institute of Metal Physics of Ural Branch of Russian Academy of Sciences, 18 S. Kovalevskaya St., Yekaterinburg 620137, Russia}
\affiliation{Skolkovo Institute of Science and Technology, 30 Bolshoy Boulevard, bld. 1, Moscow 121205, Russia}

\date{\today}

\begin{abstract}
We explore the electronic and magnetic properties of NH$_4^+$-intercalated vanadium pentoxide (NH$_4^+$-V$_2$O$_5$), a material that has been identified as a promising cathode for aqueous zinc-ion batteries. Density Functional Theory (DFT) calculations incorporating the Hubbard U correction reveal that NH$_4^+$-V$_2$O$_5$ is an antiferromagnetic insulator with an energy gap of 1.97 eV. The introduction of NH$_4^+$ ions into the V$_2$O$_5$ structure results in significant structural distortions, leading to the formation of magnetic vanadium ions (V$^{4+}$, $3d^1$) and non-magnetic ions (V$^{5+}$, $3d^0$). The magnetic vanadium ions are organized into chains with antiferromagnetic order along the $a$-axis. Our findings indicate that NH$_4^+$-V$_2$O$_5$ exhibits strong electron correlation effects and negligible interchain magnetic interactions, rendering it a promising candidate for a one-dimensional magnetic van der Waals system. This work provides insights into the electronic and magnetic behaviors of NH$_4^+$-V$_2$O$_5$, with implications for its potential applications in energy storage technologies.
\end{abstract}


\maketitle 

\section{Introduction}
Aqueous zinc-ion batteries have gained significant attention~\cite{li2024, zhao2021} for their safety, cost-effectiveness, and environmental friendliness, and V$_2$O$_5$ has emerged as a promising cathode material due to its high capacity and favorable intercalation properties~\cite{zhang2018,ding2018,yan2018}. Recent studies have shown that modifying V$_2$O$_5$ through intercalation with various ions~\cite{yang2019, tolstopyatova2022}, including ammonium~\cite{qu2021}, can enhance its structural stability and electrochemical performance, making it a strong candidate for high-performance zinc-ion batteries. 

A recent paper by Y. Wang~{\em et al.}~\cite{wang2023} proposes \nhvo with the NH$_4^+$ ions intercalated in the interlayer spacing of vanadium pentoxide as a perspective material for the cathode material for aqueous zinc-ion batteries.
It was obtained that the charge transfer, Zn-ion diffusion barrier, and the electrostatic interaction with Zn ions obviously dropped in \nhvo revealing effective accelerated electron transfer and Zn-ion migration. Authors demonstrate that the appearance of ammonia ions in the layered structure of V$_2$O$_5$ induce variations in the electronic structure of host material that result in the partial occupation of V $t_{2g}$ subshell.

The emergence of a partially filled $d$-subshell of the transition metal ion serves as a direct indicator of the existence of electron correlations effects. It could lead to electron localization on specific sites and subsequently give rise to magnetic or orbital ordering. In the present study, we explored the electronic and magnetic structure of \nhvo using the DFT+U approach, which accounts for strong electronic correlations. We demonstrated that it is energetically favored when half of the vanadium ions in the crystal cell of V$_2$O$_5$ capture electrons from ammonia ions, resulting in magnetization. \nhvo is an insulator with an energy gap of almost 2~eV. The magnetic vanadium ions form chains of antiferromagnetically ordered moments along the $a$ crystal axis.

\begin{figure}
    \centering
    \includegraphics[width=\linewidth]{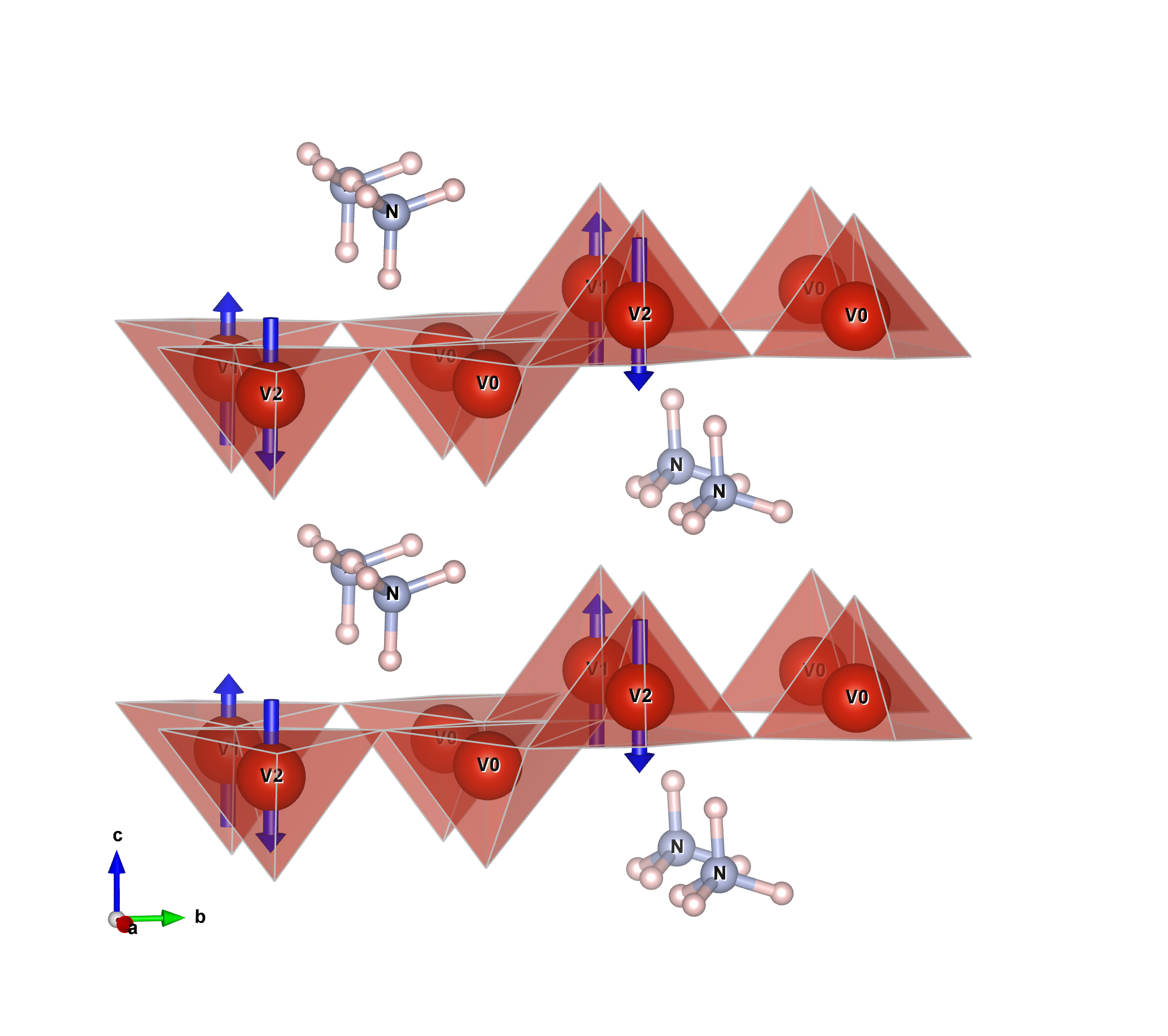}
    \caption{Relaxed crystal structure of \nhvo. Red pyramids are formed by vanadium and oxygen ions. Ammonia ions are in the interlayer space of the host vanadium pentoxide. Blue vertical arrows indicate magnetic moment orientation. Rendered using VESTA~\cite{VESTA}.}
    \label{fig:struc}
\end{figure}

\section{Methods}

All calculations were performed using the plane-wave (PW) pseudopotential implementation of the DFT contained in the Quantum ESPRESSO distribution~\cite{Giannozzi2009,giannozzi2020}.
We used pseudopotentials from the standard solid-state pseudopotential library set with the exchange-correlation functional in the PBEsol form (SSSP PBEsol Precision v1.3.0)~\cite{prandini2018}. The energy cutoff for the plane wave function and charge density expansion was set to 80~Ry and 600~Ry, respectively. The reciprocal space integration was performed on a regular $4\times2\times8$ $k$ point mesh in the irreducible part of the Brillouin zone. 

Grimme's D3 dispersion correction method~\cite{grimme2010} with three body terms was used to describe the van der Waals interactions.
The crystal structure was optimized with a convergence threshold for total energy of $< 10^{-5}$~Ry, for forces of $< 10^{-3}$~Ry/Bohr, and for pressure of $< 0.2$~kbar.
Calculations were done for the cell containing four formula units, {\em i.e.} the primitive cell doubled in the [100] direction.

The DFT+U method in the Dudarev formulation~\cite{DudarevLDA+U} with the $U_{eff}$ value of 3.2~eV~\cite{wang2023,wu2019} for the vanadium $d$ shell was used for all calculations.
The value of the Hubbard $U$ parameter was chosen based on established values from the literature for vanadium oxide systems. For example, in pristine V$_2$O$_5$, a $U$ of 3.1~eV has been employed by matching calculated and experimental oxidation enthalpies\cite{Wang2006}. For Li-intercalated vanadium pentoxide, a value of 4.0~eV is commonly used\cite{Scanlon2008, Watthaisong2019TransportPO}. While it is possible to compute $U$ self-consistently, such calculations can be sensitive to local structural parameters, as noted by Timrov et al.~\cite{timrov2022}, and are resource-intensive.

In this work, we focused on the key qualitative features of the system, including magnetic ordering, insulating behavior, and the mixed valence states of the vanadium ions. These features are largely insensitive to the precise value of $U$, as any $U$ greater than 1 eV is sufficient to open a band gap and induce magnetic ordering in the considered system due to the narrow $d_{xy}$ peak in the vanadium density of states, as demonstrated below.

\section{Results}

In stoichiometric V$_2$O$_5$, the vanadium ion has a formal electronic configuration of $d^0$. The quasi-two-dimensional crystal structure consists of edge- and corner-sharing VO$_5$ pyramids forming layers perpendicular to the (001) direction~\cite{villarsa}. If you take the crystal structure of \nhvo shown in Fig.~\ref{fig:struc} and ignore the NH$_4^+$ ions, you get a schematic view of the pure V$_2$O$_5$. In the crystal $c$-direction, the atomic layers are weakly bonded, presumably by van der Waals interactions.

In the square pyramid environment, the atomic $d$-level of V splits into four sublevels (from the lowest to the highest): $d_{xy}$, $d_{xz,yz}$, $d_{x^2-y^2}$, and $d_{3z^2-r^2}$. These sublevels are formally unoccuped. An extra electron appearing in the cell after ammonia ion intercalation will occupy the lowest level above the Fermi energy, $d_{xy}$, since the nitrogen and hydrogen energy bands are far below and above the Fermi level.

The usual GGA and LDA exchange correlation functionals lead to an averaging of occupations of all V-$d_{xy}$ orbitals in the cell due to the self-interaction error and underestimation of the Coulomb interaction effect. Consequently, the simplest DFT calculation gives a metallic electronic structure for \nhvo. If the DFT+U method is applied with spin polarization, but the symmetry of the system is not lowered, the extra electron from the NH$_4^+$ ion is again shared between the two V ions in the crystal cell, resulting in a half-filled $d_{xy}$ spin orbital and a metallic solution. The most unrestricted strategy, namely utilizing the DFT+U technique with disabled crystal symmetry and complete relaxation of atomic positions, might cause the electron to localize on just one of the two V-$d_{xy}$ orbitals. This could produce an insulating band structure and induce magnetic moments on the vanadium atoms.

We considered five different possible positions of the ammonia ions in the stoichiometric V$_2$O$_5$ (see Supplemental Materials~\cite{supplemental}). After full relaxation of the cell, the lowest enthalpy has the structure shown in Fig.~\ref{fig:struc}. The ammonia ions in \nhvo are in the $4e$ Wyckoff position of the $Pmmn$ space group of V$_2$O$_5$. 

Following this structural analysis, we calculated the formation energy of \nhvo by comparing the total energy of the fully relaxed \nhvo structure with those of the isolated NH$_4^+$ cation and pristine V$_2$O$_5$. The formation energy of -4.38~eV indicates that NH$_4^+$ incorporation into V$_2$O$_5$ is energetically favorable, suggesting a stable configuration for the ammonium-intercalated material. 
Notably, this value is more negative than the formation energy of Zn$^{2+}$ ion intercalation into pristine V$_2$O$_5$~\cite{wang2023}, further highlighting that the ammonia-intercalated structure provides a more stable host material for Zn-ion batteries.

Further insights into the structural effects of NH$_4^+$ incorporation are provided in Table~\ref{tab:lattice}, which shows the differences in cell volume and lattice parameters between V$_2$O$_5$ and \nhvo. 
The 37\% increase in cell volume is primarily driven by the expansion of the interlayer distance from 4.37~\AA to 5.98~\AA.

The obtained interlayer distance is smaller than the experimental estimate by 10.69~\AA, as reported in the study~\cite{wang2023}. We suggest that this discrepancy is due to temperature effects and possible structural defects present in the samples studied by the authors.  Nevertheless, the expansion of the interlayer spacing upon the addition of ammonium ions to vanadium pentoxide does, in fact, facilitate the potential migration of zinc ions from a purely geometric perspective.

\begin{table}
    \centering
    \begin{tabular}{c|c|c|c|c} 
    \hline \hline 
         Compound&  a, \AA&  b, \AA&  c, \AA& V$_{cell}$, \AA$^3$\\ \hline 
         V$_2$O$_5$&  7.128&  11.519&  4.373& 359.06\\ 
         \nhvo&  7.486&  10.979&  5.975& 491.04\\ 
         \hline \hline
    \end{tabular}
    \caption{The evolution of crystal cell parameters after NH$_4^+$ ions intercalation in the interlayer voids of V$_2$O$_5$.}
    \label{tab:lattice}
\end{table}

As the interlayer distance increases, structural changes in the local coordination of the vanadium ions are observed. In the primitive cell of the stoichiometric V$_2$O$_5$, all four VO$_5$ pyramids are structurally equivalent and have a polyhedral volume of 4.49~\AA$^3$. After the NH$_4^+$ ion insertion two different types of VO$_5$ pyramids appeared: one with a polyhedral volume of 4.85~\AA$^3$ (these are the pyramids around the V0 ion in Fig.~\ref{fig:struc}) and another with a polyhedral volume of 5.28~\AA$^3$ (pyramids around the V1 and V2 ions in Fig.~\ref{fig:struc}).

Even from a structural point of view, the formation of the two different types of vanadium ions is clearly seen. This finding will be supported below by an analysis of the electronic structure.

\begin{figure}
    \centering
    \includegraphics[width=\linewidth]{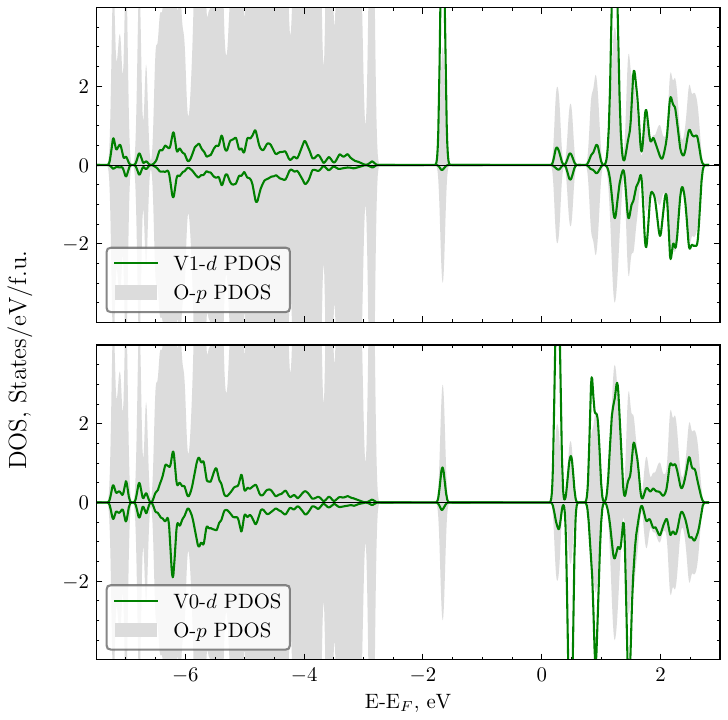}
    \caption{Spin-resolved partial densities of states for the two types of vanadium ions in \nhvo. The PDOS for O-$2p$ states is obtained by summing the contributions of all O atoms in the simulation cell. The upper part of each panel corresponds to the spin-up channel and the lower part to the spin-down channel.}
    \label{fig:pdoses}
\end{figure}

The partial densities of states for the two types of vanadium ions, called V0 and V1 in Fig.~\ref{fig:struc}, are shown in Fig.~\ref{fig:pdoses}. It can be seen that the calculation within the DFT+U approach leads to an insulating solution with an energy gap value of almost 2~eV. 
The electronic states below the -3~eV energy level are mainly formed by the $p$ states of the oxygen ions with a small mixture of V-$d$ states. The electronic states above the Fermi level are generally the V-$d$ states with an admixture of the O-$p$.
The most notable feature of the DOS is a sharp peak just below the Fermi level. This peak is mainly formed by the $d_{xy}$ orbital of the V1 ion and appears only in the spin-up channel. This indicates that the magnetic moment on the V1 ion is formed by capturing an electron from the ammonium ion and localizing it in the $d_{xy}$ orbital. In contrast, the V0 ion remains nearly non-magnetic. The partial DOS for the V2 ion is not shown, as our results indicate that V1 and V2 are structurally and electronically equivalent, with their magnetic moments arranged antiferromagnetically.

\begin{figure}
    \centering
    \includegraphics[width=0.97\linewidth]{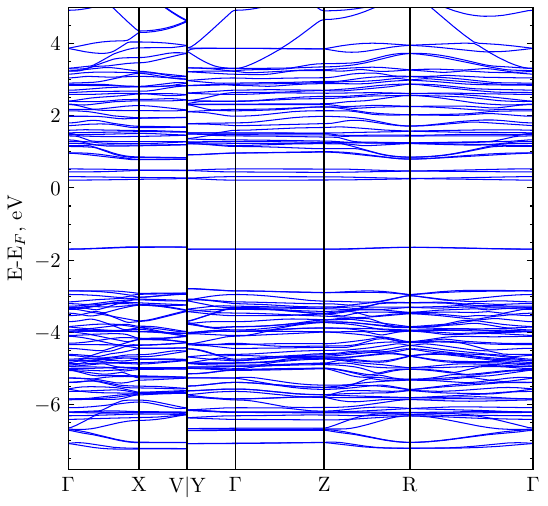}
    \caption{Band structure for the spin-up channel of NH$_4^+$-V$_2$O$_5$, calculated in the DFT+U approach. The band structure for the spin-down channel is the same, since the compound is antiferromagnetic.}
    \label{fig:bands}
\end{figure}

The calculated band structure of \nhvo, shown in Fig.~\ref{fig:bands}, reveals an ultra-narrow energy band just below the Fermi level, corresponding to the half-filled $d_{xy}$ orbitals of the V1 and V2 ions.

Finally, the lowest-energy electronic and magnetic configuration of \nhvo is as follows: after ammonia ion intercalation, half of the V ions remain in the $3d^0$ (V$^{5+}$) electronic configuration, making them non-magnetic. The other half adopt a $3d^1$, $S=\frac{1}{2}$ magnetic configuration (V$^{4+}$). These magnetic ions form linear chains along the $a$ crystal axis, separated by chains of non-magnetic V ions along the $b$ axis (see Fig.~\ref{fig:struc}). Within the chains of magnetic ions, the magnetic moments are antiferromagnetically ordered.

\begin{figure}
    \centering
    \includegraphics[width=\linewidth]{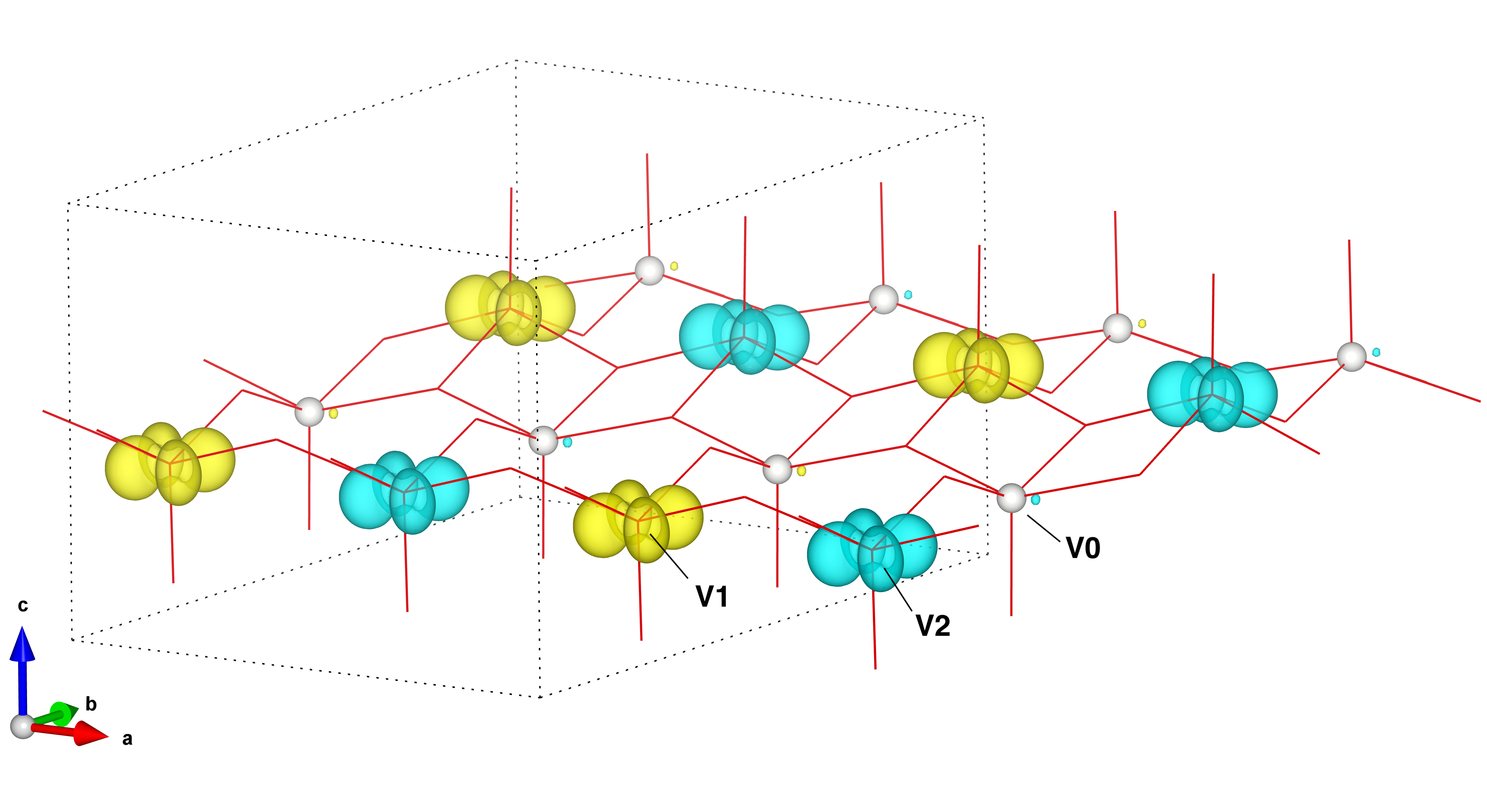}
    \caption{Magnetization density (yellow for positive and blue for negative) for \nhvo. The isosurface corresponds to 10\% of the maximum polarization value. Red lines represent vanadium-oxygen bonds. White spheres denote nonmagnetic V0 ions. The considered unit cell is shown with dotted lines. Ammonia ions are omitted for clarity.}
    \label{fig:spinrho}
\end{figure}

The arrangement of these magnetic ion chains is clearly visualized in the magnetization density isosurface ($\rho^{up} - \rho^{down}$), as shown in Fig.~\ref{fig:spinrho}. The isosurface exhibits the shape of $d_{xy}$ orbitals localized on the V1 and V2 ions. Along the $a$ axis, an alternation of magnetization sign is observed, reflecting the antiferromagnetic ordering of the magnetic moments.

According to the Goodenough-Kanamori-Anderson rules~\cite{Goodenough}, the half-filled $d$ orbitals overlapping through the $p$ orbital of the oxygen ion lying between them will form an antiferromagnetic ordering on the neighboring ions. This is the picture we see in Fig.~\ref{fig:spinrho} along the $a$ direction. To be fully confident in the result obtained, we have calculated the total energies of the cell for different magnetic moments orderings within and between the chains of magnetic ions. The corresponding total energies are summarized in Table~\ref{tab:orderings}. The obtained magnetization densities for all four possible arrangements are shown in Supplemental Materials Fig.~S2~\cite{supplemental}.

\begin{table}
    \centering
    \begin{tabular}{c|c|c} 
    \hline \hline 
         Intrachain ordering & Interchain ordering & $\Delta E$/f.u. (meV) \\
         \hline
         AFM & FM & 0 \\ 
         AFM & AFM & 0.02 \\ 
         FM & AFM & 7.08 \\ 
         FM & FM & 3.66 \\
         \hline \hline
    \end{tabular}
    \caption{The difference of the total energies between different possible arrangements between/within the chains of magnetic ions in \nhvo.}
    \label{tab:orderings}
\end{table}

The table shows that the chains of magnetic ions practically do not interact along the $b$ direction. To change the lowest interchain FM ordering to AFM would require only 0.02~meV per formula unit. 
However, 3.66~meV are required to change the magnetic ordering within the chain along the $a$ direction.
Using the mean-field approximation and neglecting interchain interactions, we
estimated the Néel temperature to be approximately 85~K.

One might expect the energy difference between parallel and antiparallel ordering of magnetic moments across chains to be minimal, assuming negligible interaction between chains, irrespective of the intrachain magnetic configuration. However, as shown in Tab.~\ref{tab:orderings}, this is not the case when the ions within the chain are ferromagnetic. The reason is that when all moments in the system (both within and between chains) are ferromagnetic, the V0 ions become more polarized. This can be observed from the magnetization density isosurface in Fig.~S2 (c) and the magnetic moments listed in Tab.~S2. The magnetic moment of the V0 ion increases from 0.12~$\mu_B$ in the AFM-intrachain FM-interchain case to 0.16~$\mu_B$ in the FM-intrachain FM-interchain case. This additional magnetic moment contributes to the total energy of the system, leading to the notable energy difference seen in the last two rows of Tab.~\ref{tab:orderings}.

\section{Conclusion}

In this study, we employed Density Functional Theory (DFT) calculations on NH$_4^+$-V$_2$O$_5$, a candidate material for a cathode in zinc-ion batteries. 
By incorporating the Hubbard U correction to account for strong electronic correlations, which increases the localization of valence electrons, and allowing for full relaxation of the crystal symmetry, we identified NH$_4^+$-V$_2$O$_5$ as an antiferromagnetic insulator with an energy gap of 1.97~eV.

Our analysis indicates that the vanadium ions in NH$_4^+$-V$_2$O$_5$ exhibit mixed valence states. Half of the vanadium ions remains in a $3d^0$ (V$^{5+}$) non-magnetic state, while the other half adopts a $3d^1$ (V$^{4+}$) configuration with localized electrons on the $d_{xy}$ orbital, forming magnetic moments. The magnetic vanadium ions are organized into linear chains along the $a$-axis with antiferromagnetic order.

It is important to note that our findings suggest negligible magnetic interactions between chains along the $b$-axis and no interaction between layers along the $c$-axis. This spatially confined magnetic interaction indicates that NH$_4^+$-V$_2$O$_5$ is a promising candidate for a one-dimensional magnetic van der Waals system, which could be advantageous for applications requiring reduced dimensionality and strong anisotropic magnetic properties.

\begin{acknowledgments}
Results were obtained with the support of the Russian Science Foundation (Project No. 24-22-00349).
\end{acknowledgments}

\bibliography{main}

\end{document}